\documentstyle[preprint,aps,epsfig]{revtex}

\begin{document}

\title{Dynamic Nonlinear Effect on Lasing in a Random Medium}

\author{B. Liu, A. Yamilov, Y. Ling, J.Y. Xu, H. Cao\footnote{h-cao@northwestern.edu}}

\address{Department of Physics and Astronomy, Northwestern University, \\ 2145 Sheridan Road, Evanston, IL 60208-3112}

%\date{\today}

\maketitle

\begin{abstract}
We have studied the dynamic effect of nonlinearity on lasing in disordered medium. The third-order nonlinearity not only changes the frequency and size of lasing modes, but also modifies the laser emission intensity and laser pulse width. When the nonlinear response time is comparable to or longer than the lifetime of the lasing mode, the nonlinearity changes the laser output through modifying the size of the lasing mode. When the nonlinear response is much faster than the buildup of the lasing mode, positive nonlinearity always extracts more laser emission from the random medium due to the enhancement of single particle scattering. 
\end{abstract}


%\begin{multicols}{2}

\newpage

Over the past decade, there have been considerable interest in nonlinear optical processes in disordered media,  e.g. second harmonic generation \cite{agraPLA88,boerPRL93,nogJOSA98}, phase conjugation \cite{kravPRB90,paasPRA97}.  It has been shown that optical nonlinearity can modify light transport in disordered medium. \cite{altJOSA86,bresEUL00} The coherent backscattering peak exhibits additional features in the presence of third-order nonlinearity. \cite{agraPRB91,heidOC95} Temporal fluctuation of scattered waves in nonlinear disordered medium leads to instability of speckle pattern. \cite{skipPRL00}

Recent studies illustrate that adding gain to a disordered medium leads to lasing in the long-lived eigenmodes. \cite{vanPRL01,burPRL01,jiaPRE02,apaPRL02,caoPRE02} Despite the modes with long lifetime are preferably amplified, their wavefunction are not modified by the presence of gain. However, nonlinearity can change the eigenmodes of a disordered system. In random lasers nonlinearity is large due to high intensity and resonance enhancement. The laser light intensity is high, especially in the case of pulsed pumping.  The nonlinearity at the lasing frequency, which is in resonance of a transition of the gain material, is resonantly enhanced. 

In this paper, we study the effect of nonlinearity on lasing in disordered medium. The experimental evidence of the change of lasing modes is presented first. It is followed  by numerical simulation of a model system. The algorithm is based on the finite-difference time-domain (FDTD) solution to Maxwell equations coupled with the rate equations of electronic populations. Third-order nonlinearity with finite response time is included. Our results demonstrate that nonlinearity not only modifies the frequency and size of the eigenmodes of a disordered system, but also changes laser emission intensity and laser pulse width of a random laser. The nonlinear response time plays a crucial role in the dynamic nonlinear effect on random lasing.  
\begin{figure}
\centering
\vspace{-0.5in}
\epsfxsize=4in \epsfbox{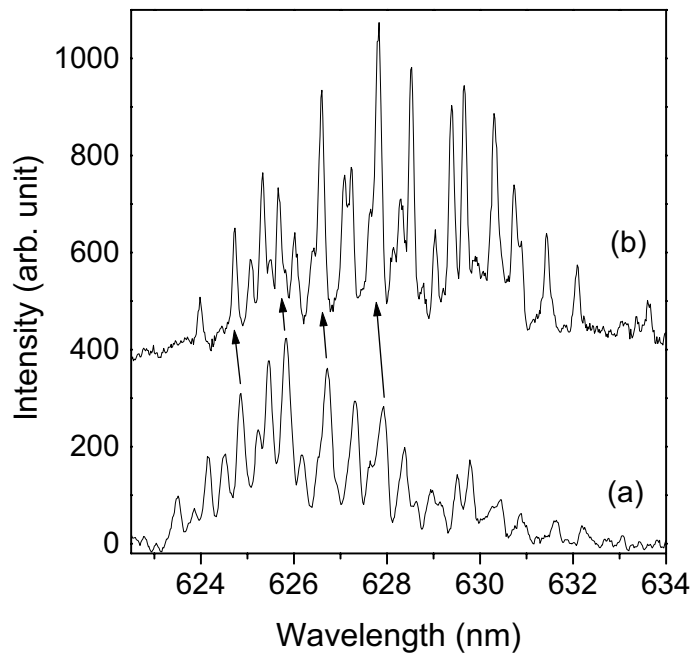}
\vspace{-0.2in}
\caption{Time-integrated spectra of laser emission from a PMMA sheet with dye and microparticles. The incident pump intensities are $2.0 \time 10^{12}$ W/m$^2$ (a), and  $4.0 \time 10^{12}$ W/m$^2$ (b).  The top spectrum is shifted vertically for clarity.}
\end{figure}
We used two types of random media in our experiments. One is poly(methyl methacrylate) (PMMA) sheets containing rhodamine 640 (Rh640) perchlorate dye and titanium dioxide (TiO$_2$)
 microparticles, the other is ZnO nanoparticles with average diameter 100 nm. The dye concentration in PMMA is 10 mM. The TiO$_2$ particles have an average diameter of
 0.4 $\mu$m. The density of TiO$_2$ particles is $\sim$ 1.4 $\times$ 10$^{12}$ cm$^{-3}$.  The PMMA sheets are optically excited by the second harmonics of a mode-locked Nd:YAG laser. The pump spot is $\sim$ 100 $\mu$m in diameter at the sample surface. The spectrum of emission from the sample is measured by a 0.5-meter spectrometer with a liquid nitrogen cooled CCD array detector. When pump intensity exceeds the lasing threshold, discrete lasing modes emerge in the emission spectrum. As the pump intensity increases, the gain spectrum of dye molecules shifts towards longer wavelength. As shown in Fig. 1, additional lasing modes appear on the long wavelength side, while some lasing modes on the short wavelength side disappear. However, individual lasing modes shift towards shorter wavelength, as marked in Fig. 1. Typically optical gain pulls the frequencies of lasing modes towards the peak of gain spectrum (frequency pulling effect). In our experiment, the lasing modes move in the opposite direction as the gain spectrum. This suggests the spectral shift of lasing modes is caused by nonlinearity. Nonlinear Kerr effect has been observed in Rh640 solution containing TiO$_2$ particles. \cite{araPRA98} The refractive index $n$ changes linearly with light intensity $I$: $n=n_0+n_2 I$, where $n_0$ is linear refractive index, $n_2$ is the nonlinear Kerr coefficient. The nonlinearity is contributed mainly by dye molecules. The value of $n_2$ in our samples is estimated to be $-10^{-14}$ to $-10^{-15}$ m$^{\rm 2}$/W. Above the lasing threshold, the excitation intensity is on the order of $10^{11}$ W/m$^{\rm 2}$. Thus the change of refractive index $\Delta n$ ranges from $-10^{-3}$ to $-10^{-4}$. This can explain the decrease of lasing wavelengths as the pump intensity increases. The wavelength shift is in the range 0.1 -- 1.0 nm, which is on the same order of the experimental shift.  

In addition to time-integrated spectra, we measured time-resolved lasing spectra to track spectral shift of lasing modes in time. Figure 2 shows a spectral-temporal image of laser emission from ZnO powder. The ZnO powder is optically excited by 20 ps pulses from a frequency-tripled mode-locked Nd:YAG laser. The pump beam is focused to a 20 $\mu$m spot on the sample surface. Laser emission from the powder is dispersed by a 0.3-meter monochromator, then directed into a Hamamatsu streak camera. Individual lasing modes are separated by the monochromator, and their temporal evolutions are recorded by the streak camera. The lasing modes at shorter wavelength are red-shifted with time, while the lasing modes at longer wavelength are blue-shifted with time. The modes in the center do not shift in wavelength. For example, the mode labeled A shifts from 387.76 nm at t = 40 ps to 387.96 nm at t = 130 ps. The mode labeled B shifts from 390.69 nm at t= 40 ps to 390.58 nm at t= 105 ps. The wavelength of mode C remains nearly constant in time. 

Temporal shift of lasing frequencies must result from dynamic change of the refractive index of ZnO. It is known that ZnO has a large third-order nonlinearity near its band edge. \cite{kalt,ravJQE92} The values of its third-order nonlinear coefficient $\chi^{(3)}$ range from $10^{-8}$ to $10^{-6}$ esu at room temperature. The nonlinear response of ZnO is also fast: its nonlinear response time is measured to be 160 fs near band edge. \cite{zhaAPL99} 

To understand the nonlinear effect on random lasing, we performed numerical simulation based on the finite-difference time-domain (FDTD) solution of Maxwell equations. \cite{taflove} The model system is one-dimensional simplification of the real experiment. The binary layers are made of dielectric materials with dielectric constants of $\epsilon_1 = \epsilon_0$ and $\epsilon_2 = 9 \epsilon_0$, respectively. The thickness of the first layer, which simulates the gain medium, is a random variable: $a = a_0 (1+ w_a \, \alpha)$, where $a_0 = 400$ nm, $w_a$ describes the strength of randomness, $\alpha$ is random value in the range [-0.5,0.5]. The thickness of the second layer, which simulates the nonlinear medium, is $b = b_0 (1+ w_b \, \beta)$, where $b_0 = 100$ nm, $w_b$ is the strength of randomness, $\beta$ is random value in the range [-0.5,0.5]. For simplicity, the gain component and the nonlinear component are separated in our model. This separation does not change qualitatively the nonlinear effects that we study.  The dynamics of electronic populations in the gain layers is described by four-level rate equations, with parameters similar to those in Ref. \cite{souPRB02} In the nonlinear layers, the third-order nonlinearity is described by the nonlinear polarization \cite{zioJOSA93,josTAP97}
\begin{equation}
P_{NL}(x,t) = \epsilon_0 \chi^{(3)} E(x,t) \int_{- \infty}^{\infty} g(t-\tau) |E(x,\tau)|^2 d \tau
\end{equation}  
$\chi^{(3)}$ is the third-order nonlinear coefficient. The casual response function $g(t-\tau)= (1/ \tau_0) \exp[-(t-\tau)/\tau_0]$ for $t \geq \tau$, and $g(t-\tau)=0$ for $t < \tau$. $\tau_0$ is the nonlinear response time. To incorporate the nonlinearity with finite response time into the FDTD algorithm, we introduce a new function 
\begin{equation}
G(x,t) \equiv \int_{- \infty}^{\infty} g(t-\tau) |E(x,\tau)|^2 d \tau = {1 \over{\tau_0}} \int_0^t e^{-(t-\tau)/\tau_0} |E(x,\tau)|^2 d \tau. 
\end{equation}
The differential equation for $G(x,t)$ can be derived as
\begin{equation}
{ {d G(x,t)} \over {dt}} =  - {{G(x,t)} \over {\tau_0}} +  {{|E(x,t)|^2} \over {\tau_0}}
\end{equation}
$E$ is related to $D$ by accounting for both linear and nonlinear polarizations: 
\begin{equation}
E(x,t) = {{D(x,t)- P_L(x,t)- P_{NL}(x,t)} \over {\epsilon_0 \epsilon_{\infty}}}
\end{equation}
According to Eq.(4), The effective dielectric constant is
\begin{equation} 
\epsilon(x,t)=\epsilon_{\infty}+ \chi^{(1)}+ \chi^{(3)} G(x,t).
\end{equation}
When light frequency is far from any resonant frequencies, $\tau_0 \rightarrow 0$, the nonlinearity becomes instantaneous.  $g(t-\tau) = \delta(t-\tau)$, $G(x,t)=|E(x,t)|^2$. The nonlinear dielectric constant $\epsilon=\epsilon_L+\chi^{(3)} |E|^2$, where $\epsilon_L = \epsilon_{\infty} + \chi^{(1)}$ is the linear dielectric constant.  
\begin{figure}
\centering
\vspace{0.1in}
\epsfxsize=4in \epsfbox{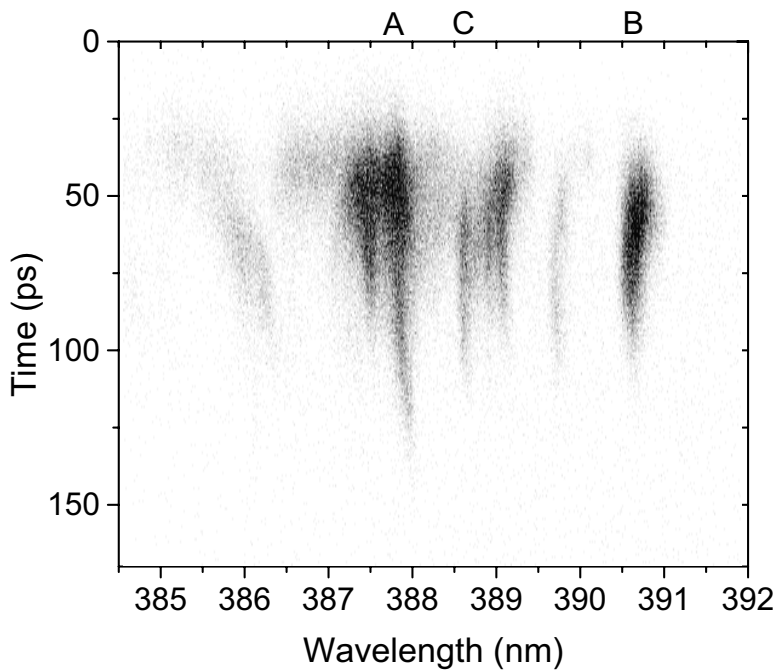}
\vspace{-0.1in}
\caption{A spectral-temporal image of laser emission from ZnO powder.}
\end{figure}
To simulate pulsed pumping, the external pumping term in the rate equations represents a Gaussian pulse of width 20 ps: $P_r(t)= 2 \times 10^8 \exp[-{\rm ln}2 \, (t-t_0)^2/t_0^2]$ 1/s, where $t_0$= 10 ps. We first turn off the nonlinearity by setting $\chi^{(3)}$ = 0 to simulate lasing in linear random medium. The random structure consists of 50 layers with $w_a = w_b = 0.9$. The  radiative transition of the four-level system is centered at $\lambda_a$ = 389 nm with 7 nm spectral width. We choose the pumping rate so that only one mode lases. The frequency of the lasing mode does not exhibit any noticeable change during the lasing process. Next we include the nonlinearity. We choose the value of  $\chi^{(3)}$ to be $1.2 \times 10^{-16}$ m$^{\rm 2}$/V$^{\rm 2}$, close to the measured value for ZnO at room temperature. \cite{zhaAPL99} Figure 3 plots temporal evolution of lasing spectra obtained by Fourier transform of the electric field in 2.6 ps intervals. For negative (positive) nonlinearity $\chi^{(3)} < 0$ ($\chi^{(3)} > 0$), the lasing mode first shifts towards shorter (longer) wavelength as the laser intensity increases, then shifts to longer (shorter) wavelength as the laser intensity decreases. These results can explain qualitatively the experimental observation of lasing frequency shift with time. As shown in Fig. 2, the laser pulses in ZnO powder rise quickly then decay slowly. The rising edge of the laser pulses is too short to detect temporal shift of the lasing frequency. However, the falling edge of the laser pulses is long enough to observe noticeable shift of lasing frequency with time. In the presence of population inversion, $\chi^{(3)}$ is positive (negative) for photon energy below (above) the band gap. For the lasing modes whose frequencies are below (above) ZnO band edge, since $\chi^{(3)} > 0$ ($\chi^{(3)} < 0$),  they shift towards shorter (longer) wavelength in the falling edge as in Fig. 3b (3a). For the lasing modes at the band gap, $\chi^{(3)} \approx 0$, thus their frequencies remain nearly constant with time. The wavelength shift in our calculation is $\sim$ 1--2 $\AA$, close to the experimental value. 
\begin{figure}
\centering
\vspace{0in}
\epsfxsize=4in \epsfbox{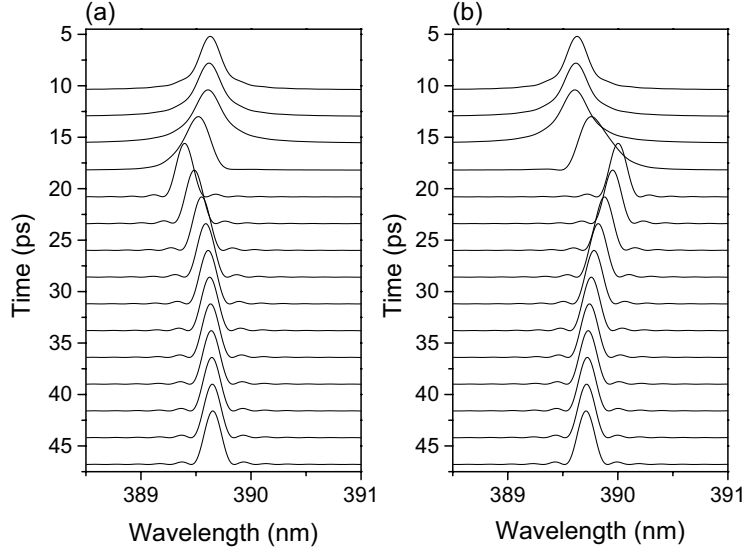}
\vspace{0in}
\caption{ Temporal evolution of lasing spectra calculated for (a) $\chi^{(3)} = -1.2 \times 10^{-16}$ m$^{\rm 2}$/V$^{\rm 2}$, (b) $\chi^{(3)} = 1.2 \times 10^{-16}$ m$^{\rm 2}$/V$^{\rm 2}$.}
\end{figure} 

The spatial size of lasing modes also changes in the presence of third-order nonlinearity. When only one mode lases, the size of the lasing mode can be characterized by the inverse partition ratio $r(t) = (\int |E(x,t)|^2 d x)^2 / \int |E(x,t)|^4 dx$. Figure 4 plots $r(t)$ of one lasing mode (the same one as in Fig. 3) in both linear and nonlinear cases. For this mode, when $\chi^{(3)} > 0$ ($\chi^{(3)} < 0$), $r(t)$ increases (decreases) during the lasing period. We repeat the calculation for lasing modes in many random configurations. However, we do not find any definite relation between the sign of the 
nonlinearity and the size change of the lasing modes. Nevertheless, in 
all simulations, we do observe a noticeable alteration in the mode size.
\begin{figure}
\centering
\vspace{-0.3in}
\epsfxsize=4in \epsfbox{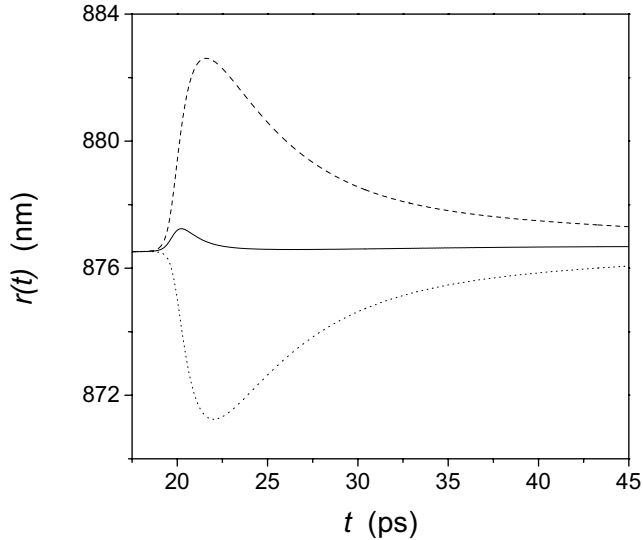}
\vspace{-0.5in}
\caption{Spatial size of the lasing mode $r(t)$ calculated for $\chi^{(3)} = 0$ (solid line), $\chi^{(3)} = -1.2 \times 10^{-16}$ m$^{\rm 2}$/V$^{\rm 2}$ (dotted line), and $\chi^{(3)} = 1.2 \times 10^{-16}$ m$^{\rm 2}$/V$^{\rm 2}$ (dashed line).}
\end{figure}
Optical nonlinearity not only changes the frequency and size of lasing modes, but also  modifies the intensity and duration of laser pulses. The change of laser output is sensitive to the nonlinear response time $\tau_0$. We calculate the laser pulses by fixing the value of $\chi^{(3)}$ and varying $\tau_0$. Figure 5 plots the total laser emission energy $U(t) = (1/2) \int \epsilon_0 \epsilon |E(x,t)|^2 dx$ for $\chi^{(3)} = \pm 1.2 \times 10^{-16}$ m$^{\rm 2}$/V$^{\rm 2}$.  We first set $\tau_0$ equal to the experimental value (160 fs) of passive ZnO near band edge \cite{zhaAPL99}. For positive (negative) $\chi^{(3)}$, the laser pulse becomes weaker (stronger) than that in the linear case. Next we repeat the calculation with smaller $\tau_0$ since the nonlinear response time is shortened in the presence of stimulated emission. As shown in Fig. 5, when $\tau_0$ = 6.5 fs, the behavior of laser pulse becomes totally different. For positive (negative) $\chi^{(3)}$, the laser pulse becomes stronger (weaker) and longer (shorter) than that in the linear case. 
\begin{figure}
\centering
\vspace{-0.3in}
\epsfxsize=4in \epsfbox{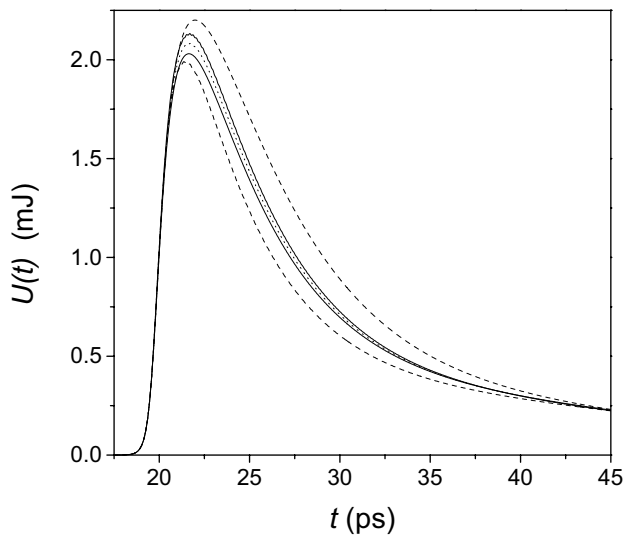}
\vspace{-0.5in}
\caption{Laser emission energy $U(t)$ calculated for (from top to bottom)  $\chi^{(3)} = 1.2 \times 10^{-16}$ m$^{\rm 2}$/V$^{\rm 2}$ and $\tau_0 = 6.5$ fs; $\chi^{(3)} = -1.2 \times 10^{-16}$ m$^{\rm 2}$/V$^{\rm 2}$ and $\tau_0 = 160$ fs; $\chi^{(3)} = 0$; $\chi^{(3)} = 1.2 \times 10^{-16}$ m$^{\rm 2}$/V$^{\rm 2}$ and $\tau_0 = 160$ fs; $\chi^{(3)} = -1.2 \times 10^{-16}$ m$^{\rm 2}$/V$^{\rm 2}$ and $\tau_0 = 6.5$ fs.}
\end{figure}
From the calculation of many random lasing modes, we conclude that the effect of nonlinearity on laser pulse intensity and width depends on the relative magnitude of two time scales. One is the nonlinear response time $\tau_0$, the other is the lifetime $\tau_c$ of the lasing mode in the passive medium. $\tau_c$ can be obtained by launching a pulse at the frequency of the lasing mode in the passive medium and observing the temporal decay of the electric field. \cite{vanPRL01} When $\tau_0$ is comparable to or longer than $\tau_c$, the change of laser pulse intensity and width is related to the size change of the lasing mode. If the size of the lasing mode decreases (increases) in the presence of nonlinearity, light confinement gets better (worse). The decrease (increase) of light leakage is equivalent to an increase (decrease) of the quality factor of the random cavity. Hence, lasing lasts longer (shorter), and laser emission is stronger (weaker). When $\tau_0$ is much shorter than $\tau_c$, the change of laser output depends only on the sign of $\chi^{(3)}$, i.e., positive (negative) nonlinearity always extract more (less) laser emission from the random medium at the same pumping rate. When the nonlinear response is much faster than the buildup of the lasing mode, the lasing mode cannot response fast enough to the nonlinear refractive index change. The phase of scattered light changes quickly due to rapid change of refractive index with intensity. The absence of constant phase relations among light waves scattered by different particles undermines the interference effect. Hence, the effect of single particle scattering becomes dominant over the collective effect of many particle scattering. For $\chi^{(3)} > 0$ ($\chi^{(3)} < 0$), the refractive index contrast of the binary layers increases (decreases) as the laser intensity increases. Light scattering of a single particle becomes stronger (weaker). The increase (decrease) of scattering strength results in more (less) efficient lasing, i.e., higher (lower) laser intensity and longer (shorter) lasing period. 

In conclusion, we have studied the effect of nonlinearity on lasing in disordered medium. The third-order nonlinearity not only changes the frequency and size of lasing modes, but also modifies the laser emission intensity and laser pulse duration. When the nonlinear response time is comparable to or longer than the lifetime of the lasing mode, the nonlinearity changes the laser output through modifying the size of the lasing mode or equivalently the quality factor of the random cavity. When the nonlinear response is much faster than the buildup of the lasing mode, positive nonlinearity always extracts more laser emission from the random medium due to the enhancement of single particle scattering. 

We acknowledge Dr. A.L. Burin and S.-H. Chang for stimulating discussions. This work is supported partially by the National Science Foundation under Grant No. DMR-0093949 and the David and Lucille Packard Foundation.

\end{document}